\documentclass[5p,twocolumn]{elsarticle}
\usepackage{graphicx}
\usepackage{graphicx}
\usepackage{amsmath,amssymb}
\newcommand\beq{\begin{equation}}
\newcommand\eeq{\end{equation}}
\newcommand\bea{\begin{eqnarray}}
\newcommand\eea{\end{eqnarray}}
\usepackage{graphicx}
\usepackage{graphicx}
\usepackage{xcolor}
\begin{document}
\title{\Large \bf From Lattice Gauge Theories to Hydrogen Atoms}
\author[snb]{Manu Mathur}
\ead{manu@bose.res.in}
\author[snb]{T. P. Sreeraj}
\ead{sreerajtp@bose.res.in}
\address[snb]{S. N. Bose National Centre for Basic Sciences, Salt Lake, JD Block, Sector 3, Kolkata 700098, India}

\begin{abstract}
\noindent We construct canonical transformations 
to obtain a complete and most economical realization of the 
physical Hilbert space ${\cal H}^p$ of pure $SU(2)_{2+1}$ lattice gauge theory in terms of Wigner coupled 
Hilbert spaces of hydrogen atoms. One hydrogen atom is assigned to every plaquette of the lattice. 
A complete orthonormal description of the Wilson loop basis in ${\cal H}^p$ is obtained by all possible angular momentum Wigner couplings of hydrogen atom energy eigenstates $\vert n~l~m\rangle$ describing electric fluxes on the loops. The SU(2) gauge invariance implies that the total angular momenta of all hydrogen atoms vanish.
The canonical transformations also enable us to rewrite
the  Kogut-Susskind  Hamiltonian in terms of  fundamental Wilson loop operators and their 
conjugate electric fields. The resulting loop Hamiltonian 
has  a global SU(2) invariance and a simple 
weak coupling ($g^2\rightarrow 0$) continuum  limit. 
The canonical transformations leading to the loop Hamiltonian are valid for any SU(N). The  ideas and techniques 
can also be extended to higher dimension.  
\end{abstract}

\maketitle

\section{Introduction}
\label{sintro} 

\noindent The idea that gauge  theories should be reformulated as theory of loops and strings without any color degrees of freedom is old  
\cite{mans1} and refuses to die. One hopes that eventually  
some appropriate loop  description of gauge theories will provide a natural and elegant framework to compute low energy QCD effects
leading to a better understanding of non-perturbative issues like color confinement. Since the work of Ashtekar, loops carrying SU(2) 
fluxes  have also found their relevance in quantum gravity  where they  
describe quantum excitations of geometry 
\cite{rov}. In condensed matter physics \cite{cmp}, many effective models are in 
terms of non-abelian gauge theories. In view of above, the importance of developing new ideas and techniques to understand gauge 
theories better requires no emphasis. In recent past, quest 
to realize simplest non-abelian SU(2) lattice gauge theory 
Hamiltonian dynamics using cold atomic gases in optical lattices \cite{cirac} as well as  SU(2) tensor network ideas \cite{tna} to explore the interesting part of the physical Hilbert space ${\cal H}^p$ are  important and exciting new developments in this direction. 

\begin{figure}[]
\begin{center}
\includegraphics[width=0.34\textwidth,height=0.34\textwidth]
{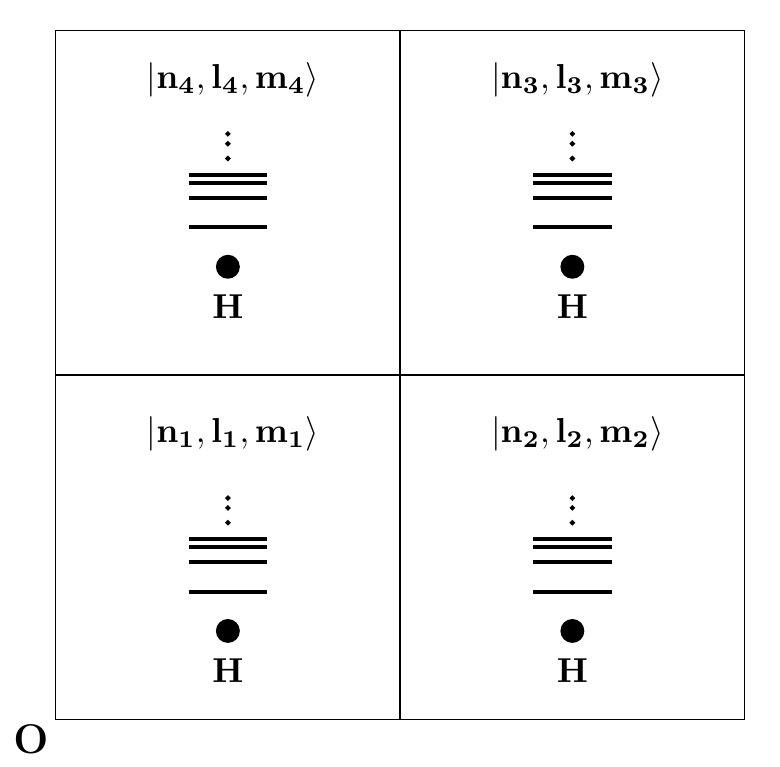} 
\end{center}
\caption{One Hydrogen atom, denoted by $\bullet$, is assigned to  each plaquette. The energy eigenstates $|{n_p~l_p~m_p}\rangle$ 
Wigner couple together (\ref{entlat})  with 
vanishing total angular momenta to form a basis in
the physical Hilbert space  ${\cal H}^p$ of pure SU(2) lattice gauge theory.}
\label{ha}
\end{figure}

In this letter, 
we show that 
the gauge invariant physical or loop Hilbert space 
${\cal H}^p$ 
of pure SU(2) lattice gauge theory  
can be completely and most economically realized in terms of the Wigner coupled bound energy eigenstates $\vert n~l~m\rangle$ of hydrogen atoms
associated with the plaquettes of the lattice as shown in figure (\ref{ha}).  This Wigner coupled hydrogen atom basis describes  quantized SU(2) loop electric fluxes in terms of $(n,l,m)$ and is orthonormal as well as complete in ${\cal H}^p$. Therefore, our description of loop dynamics  is completely  free of the notorious Mandelstam constraints\footnote{In fact, these constraints have been major obstacles in the loop formulation because of  their non-local nature \cite{manu,loops}. They have been extensively discussed in the past \cite{loops} in the context of SU(2) gauge theory leading to  spin networks as  their solutions \cite{manu,sharat,loops}. However, loop dynamics is extremely complicated in the spin network basis \cite{manu,sharat}.}. We rewrite the Kogut-Susskind Hamiltonian in terms of the fundamental plaquette loop operators and conjugate loop electric fields.  We show that this loop Hamiltonian has a global SU(2) symmetry and there are no gauge fields. Further, the loop Hamiltonian has a simple weak coupling ($g^2\rightarrow0$) limit. 
We work in two space dimension on a finite lattice with ${\cal N} (=(N+1)\times (N+1))$ sites, ${\cal L}$ 
links and ${\cal P}$ plaquettes satisfying:
${\cal L} ={\cal P} + \left({\cal N}-1\right)$. 
We choose open boundary conditions. A lattice site is denoted by $(x,y)$ with $x,y=0,1,\cdots ,N$. There are 
no static or dynamical charges. All  ``hydrogen atom solutions" of the local Gauss law constraints  are obtained in the  charge zero sector.  

Following Fock \cite{pauli}, we  describe ${\cal P}$ hydrogen atoms in figure (\ref{ha}) on their momentum hypersphere $S^3$ so that their hidden $SU(2) \times SU(2) $ symmetries become manifest (see section (\ref{sha}), equation (\ref{transhyd})). 
On the other hand, we construct SU(2) (SU(N)) canonical transformations which fuse ${\cal L}$  Kogut-Susskind link operators  into ${\cal P}$ fundamental plaquette loop operators, each enclosing one of the ${\cal P}$ plaquettes
as shown in figure (\ref{f:4}-a). 
The equivalence of the  gauge theory and hydrogen atom  Hilbert spaces has its origin in the identification of  SU(2) group manifold $S^3$ associated with each plaquette loop holonomy  with the $S^3$ of the corresponding hydrogen atom.     
The plan of the paper is as follows.  In section (\ref{sha})  we  briefly review hydrogen atom Hilbert space  in a language  \cite{wyb} which helps us later  in establishing  its connection with lattice gauge  theory. The section (\ref{sct}) discusses the canonical transformations which fuse all Kogut-Susskind SU(N) link operators into mutually independent and fundamental Wilson loop operators. Having discussed 
hydrogen atoms and loop formulation separately,  in section (\ref{slsha}) we establish the exact equivalence of the Hilbert spaces of hydrogen atoms and SU(2) 
lattice gauge theory. In  section (\ref{sldlo}) we 
construct loop Hamiltonian and briefly discuss loop dynamics. We end the paper with a short discussion on 
our results and some of the possible future directions.  
\section{Hydrogen Atom}
\label{sha}
As shown by Pauli \cite{pauli} in 1926, the hydrogen atom can be elegantly solved using group theory which exploits manifest rotational and hidden Runge Lenz symmetries generated by angular momentum ($\vec{L}$) and Laplace Runge Lenz ($\vec{A}$) operators\footnote{We follow Wybourne \cite{wyb} for hydrogen atom discussions. The Runge Lenz vector has been scaled by by $\frac{1}{\sqrt{-2H}}$.} respectively. 
These generators commute with the hydrogen atom Hamiltonian and satisfy $\vec{L}\cdot \vec{A} = 0$. 
This leads to $SU(2)\otimes SU(2)$ symmetry algebra generated by  $\vec{J}_{\pm} \equiv {1}/{2} (\vec{L}\pm \vec{A})$ 
on the bound states of hydrogen atom $(E < 0)$ \cite{wyb}. 
Further, as $\left[J^a_+,J^b_-\right] = 0$ and $\vec{J}_+^{~2} = \vec{J}_-^{~2} \equiv \vec{J}^{~2}$,   
the two equivalent complete set of commuting operators (CSCO) are $[\vec{J}^{~2}, J_+^z, J^z_- ]$ (CSCO-I) in the decoupled basis and  
$[\vec{J}^{~2},~\vec{L}^2,~\vec{L}_z]$ (CSCO-II) in the coupled basis. Following Wybourne \cite{wyb}, we  define\footnote{$J^a_{\mp}$ should not be confused with the raising and lowering angular momentum operators.}: 
\begin{align}
J_-^a \equiv a^\dagger \left(\frac{\sigma^a}{2}\right) a;~~  J_+^a \equiv b^\dagger \left(\frac{\sigma^a}{2}\right) b. 
\label{hydsb} 
\end{align}
In (\ref{hydsb}) $(a^\dagger_1,a^\dagger_2)$ and $(b^\dagger_1,b^\dagger_2)$ represent SU(2) doublets of Schwinger boson creation operators,
$\sigma^a$ (a=1,2,3) are the  
Pauli matrices. 
The condition $\vec J_-^{~2} = \vec J_+^{~2}$ implies $N_a = N_b $ where $N_a= a^\dagger \cdot a$ and $N_b  = b^\dagger \cdot b$ 
are the total number operators. The orthonormal and complete basis diagonalizing CSCO-I is given by \cite{wyb}: 
\bea 
\label{csco1} 
|j=j_-=j_+,m_-, m_+ \rangle  =  |j,m_-\rangle \otimes |j, m_+\rangle 
\eea 
\bea 
|j_-=j,m_-\rangle \equiv   \frac{(a^{\dagger}_1)^{(j+m_-)} (a^{\dagger}_2)^{(j-m_-)}}{(j+m_-)!(j-m_-)!} |0\rangle \nonumber \\    
|j_+=j, m_+\rangle  \equiv  \frac{(b^{\dagger}_1)^{(j+ m_+)} (b^{\dagger}_2)^{(j- m_+)}}{(j+ m_+)!(j- m_+)!} |0\rangle. \nonumber      
\eea 
The other equivalent coupled hydrogen atom basis diagonalizing the CSCO-II is given by: 
\bea 
\vert n~l~m \rangle & \equiv & \sum_{m_-, m_+} C_{jm_-,j m_+}^{~~l,m}~ |j,m_-, m_+\rangle 
\label{nlm} 
\eea 
In (\ref{nlm}), $n \equiv (2j+1) =1,2\cdots;l=0,1,2,\cdots,(n-1);m =-l,\cdots,+l$; $C_{jm^\prime,j\bar m}^{~~l,m}$ are 
the Clebsch-Gordan coefficients. The hydrogen atom states $|n~l~m\rangle$ are 
eigenstates of  $J^2, L^2, L_z$ and also of the Hamiltonian  with energy \cite{wyb} $E \sim -1/n^2$. 
For later purpose, it is convenient to graphically  represent 
the hydrogen atom states $|n~l~m\rangle$ in (\ref{nlm}) by a tadpole in figure (\ref{fama}).  

As shown by Fock, the above $SU(2) \otimes SU(2)$ symmetry for bound states ($p_0^2\equiv -2E >0$) becomes manifest
if we transcribe 
the hydrogen atom dynamics on a hypersphere $S^3 :(q_0,~\vec{q};~ q_0^2+\vec{q}^{~2}=1)$ embedded in 
$R^4:(p_0,p_1,p_2,p_3)$ 
through a stereographic projection: 
\bea 
\label{transhyd} 
&&{q}_0  \equiv  \frac{(p_0^2-\vec{p}^{~2)}}{(p_0^2+\vec{p}^{~2})}, ~~ \vec{q}  \equiv  \frac{2p_0\vec{p}}{(p_0^2+\vec{p}^{~2})}, \\
&&\Omega_H(q_0,\vec q)\equiv q_0 \sigma_0 + i \vec{q} \cdot \vec \sigma,  ~~~  q_0^2 + \vec{q}^{~2} =1. 
\nonumber
\eea 
Above $\sigma_0,\vec \sigma$ are the identity, Pauli matrices respectively. 
The mapping (\ref{transhyd}) enables us to transform 
\cite{pauli}  momentum space hydrogen atom Schrodinger 
equation into the integral  equation  of the  4-dimensional spherical  harmonics $Y_{n,l,m}(\Omega_H)$
representing a free particle on $S^3$.
It was later shown by  Bargmann \cite{pauli} that 
$(L_1,L_2,L_3)$ and $(A_1,A_2,A_3)$ correspond to rotations in $(q_2q_3), (q_1q_3), (q_1q_2)$ and 
$(q_0q_1), (q_0q_2), (q_0q_3)$ planes respectively making   $SU(2)\otimes SU(2)$ symmetry of hydrogen atom  manifest.

We  now construct iterative canonical transformations which make all gauge degrees of freedom decouple from ${\cal H}^p$ by systematically fusing the Kogut-Susskind \cite{kogut} link flux operators into a complete and mutually independent set of plaquette loop operators. These loop operators  (\ref{holo}),(\ref{holop}) in turn are then associated with hydrogen atoms. 
To keep the presentation simple and short, we always illustrate the ideas and techniques on a single plaquette and then  generalize the results to the entire lattice.
We have chosen space dimension $d=2$ for the same reasons. 

\section{Canonical Transformations: From Links to Loops} 
\label{sct} 
\subsection{Single Plaquette Case}

\begin{figure}[]
\begin{center}
\includegraphics[scale=1.1]
{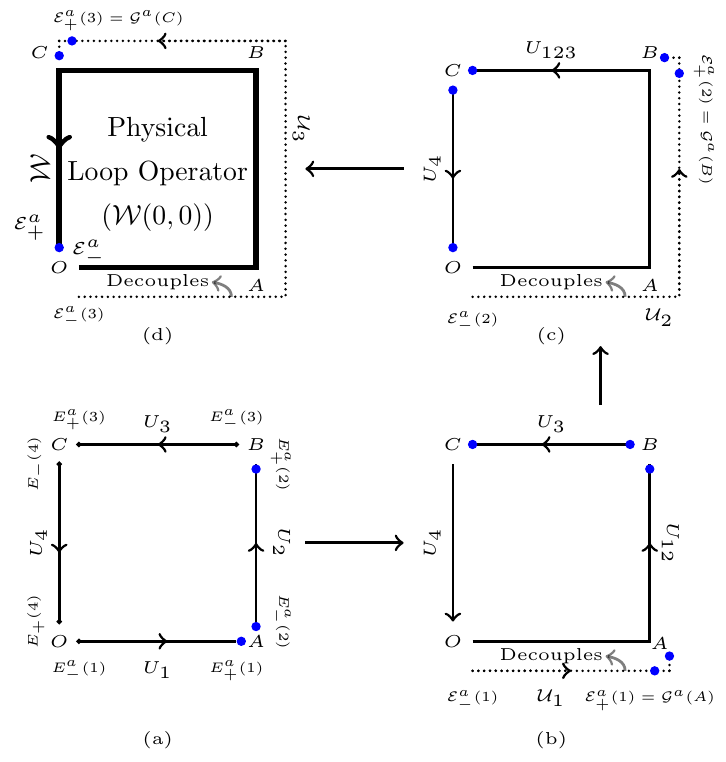} 
\end{center}
\caption{Three canonical transformations on the four link flux operators of a plaquette leading to a single plaquette loop flux operator ${\cal W}_{\alpha\beta}$. The electric fields involved in the canonical transformations are represented by {\tiny{ \textcolor{blue}{$\bullet$}}} .Two {\tiny{ \textcolor{blue}{$\bullet$}}} at the end of flux operators ${\cal U}(1), {\cal U}(2)$ and 
${\cal U}(3)$ represent their electric fields which are also the Gauss law operators ${\cal G}^a$ at A, B and C respectively. Only the loop operator $\left({\cal W}_{\alpha\beta}, ~ {\cal E}_\mp\right)$ are relevant, the remaining are unphysical and to be ignored.} 
\label{f:2}
\end{figure}
We start with a plaquette OABC with 4 Kogut-Susskind SU(2) flux operators \cite{kogut} $U_I, I =1,2,3,4$ attached to 4 sides. The left, right rotations on $U_l$ are generated by the left, right electric fields 
$E^a_-(l), ~E^a_+(l)$ (a =1,2,3) respectively as shown in 
figure (\ref{f:2}-a).  The  basic quantization rules are:  
\bea 
 \left[E^a_+,U_{\alpha\beta}\right] = \left(U \frac{\sigma^a}{2}\right)_{\alpha\beta} =>   
\left[E^a_+,E^b_+\right] = i \epsilon^{abc} E^c_+ \nonumber \\  
\hspace{-0.9cm} \left[E^a_-,U_{\alpha\beta}\right] = - \left(\frac{\sigma^a}{2} U\right)_{\alpha\beta}  =>   
\left[E^a_-,E^b_-\right] = i \epsilon^{abc} E^c_-      
\label{ccr}
\eea 
As operators on different links commute, we have suppressed the link index $l$ in (\ref{ccr}). 
One can check that $E_+^a = - R_{ab}(U^\dagger) E_-^b$ where $R_{ab}(U) = \frac{1}{2} Tr\left(\sigma^a U \sigma^bU^\dagger\right)$ is 
SO(3) rotation matrix implying: 
$(\vec{E}_-)^2 = (\vec{E}_+)^2, ~~\left[E^a_-,E^b_+\right] = 0,$
on every links. 

We now make canonical transformations to fuse 
$\{U_1, U_2, U_3, U_4\}$
into 3 unphysical string flux operators $\{{\cal U}_1, {\cal U}_2, {\cal U}_3\}$  
and a physical Wilson plaquette loop  operator\footnote{Note that the shapes of loops and strings will depend on the canonical transformations.} ${\cal W}$ around OABC as shown in figure \ref{f:2}. The corresponding left, right string and loop
electric fields are denoted by ${\cal E}_{\mp}^a(I)~(I=1,2,3)$ and ${\cal E}_{\mp}^a$ respectively as shown in figure (\ref{f:2}-a,b,c,d). 
The first canonical transformation fusing $U_1$ and $U_2$,  is:
\begin{eqnarray}
\label{ct1}  
U_1 \rightarrow {\cal U}_1  \equiv   U_1,  ~~~U_2 \rightarrow {{U}}_{12} \equiv U_1 U_2,  ~~~~~~~ \\ 
  =>  {\cal{E}}^a_+(1)  =  {E^a_+(1) + {E}^a_-(2)}, ~~~{{E}}^a_+(12)  = {E}^a_+(2).      \nonumber 
\end{eqnarray} 
Note that the new pairs $\left\{{\cal U}(1), {\cal E}_+(1)\right\}$ and $\left\{{U}_{12}, {E}_+(12)\right\}$ are canonical satisfying (\ref{ccr}) and also mutually independent.  The 
SU(2) Gauss law
 at the corner A in figure (\ref{f:2}) states\footnote{
  $A\approxeq 0$ implies that the operator A annihilates all physical states  in ${\cal H}^p$.}, 
${\mathcal G}^a(A)={\cal E}^a_+(1) = E_+^a(1) + E_-^a(2) \approxeq 0$ in ${\cal H}^p$.  Hence,  ${\cal U}_1$ completely 
decouples from ${\cal H}^p$. We now iterate the canonical transformation 
(\ref{ct1}) with $U_1, U_2$ replaced by ${U}_{12}, U_3$: 
\beq
\label{ct2}  
 {\cal U}(2)  \equiv  {U}_{12}, ~~~{U}_{123} \equiv {U}_{12} U_3, 
 \eeq  
 \beq
=>  {\cal E}^a_+(2)  =  {E}^a_+(12) + {E}^a_-(3),  ~~  {E}^a_+(123)  = {E}^a_+(3). \nonumber      
\eeq 
We  get ${\cal G}^a(B)={\cal E}^a_+(2)  = {E}^a_+(12) + {E}^a_-(3)= E^a_+(2)+E^a_-(3) \approxeq 0$ (Gauss law at B) and  ${\cal U}_2$ decouples.  
The last canonical transformation on the plaquette OABC  is: 
\begin{eqnarray} 
\label{ct3} 
{\cal U}(3)  \equiv  U_{123},~~~~~~~~~~~~~~~~~ {\cal W} \equiv  U_{123} U(4),  \\ 
{\cal E}^a_+(3)  =  E^a_+(123) + {E}^a_-(4) 
\approxeq 0, ~~~~~ {\cal E}^a_+  = {E}^a_+(4). \nonumber       
\end{eqnarray} 
Now ${\cal U}_3$ decouples as ${\cal G}^a(C)={\cal E}^a_+(3)$. 
Therefore, we need to focus only on the plaquette loop operators $\left\{{\cal W}, {\cal E}_{\pm}^a\right\}$.  
They are covariant under gauge transformation 
(\ref{gtl}) at O. The SU(2) Gauss laws at A, B, C remove 
the strings ${\cal U}_1, {\cal U}_2$ and ${\cal U}_3$ respectively.  The Gauss law at the origin, corresponding to the covariant transformations of the physical loop operators: 
\bea 
 {\cal E}_{\mp} \rightarrow \Lambda_0 ~{\cal E}_{\mp}~ \Lambda^\dagger_0,~~~~ ~~~~{\cal W}  \rightarrow \Lambda_0 ~~{\cal W} ~ \Lambda^\dagger_0,  
\label{gtlp}
\eea 
reduces to  global constraints:
 \begin{align}  
{\mathcal G}^a={\sf L}^a  \equiv   {\cal E}_-^a + {\cal E}_+^a =0.
\label{gl}
\end{align}
It is convenient to describe SU(2) loop electric field and flux operators in 
terms of the prepotential doublets \cite{manu}: 
\begin{align}
{\cal E}_-^a = (1/2)~ a^\dagger ~\sigma^a ~a~~~~~~{\cal E}_+^a = (1/2)~ b^\dagger ~{\sigma^a}~  b \nonumber \\ 
{\cal W}_{\alpha\beta} =  \frac{1}{\sqrt{(N+1)}} \Big(~\tilde a^{\dagger}_{\alpha}~  
b^{\dagger}_{\beta}~  - ~a_{\alpha} ~\tilde b_{\beta}~\Big) 
\frac{1}{\sqrt{(N+1)}}.    
\label{holo}
\end{align}
Above ${\tilde a}_\alpha \equiv \epsilon_{\alpha\gamma} a_\gamma$.  Like link electric fields \cite{kogut}, loop electric fields satisfy $(\vec{\cal E}_-)^2 = (\vec{\cal E}_+)^2$  
implying $N_a=N_b \equiv N$. 



The  transformations (\ref{ct1}), (\ref{ct2}) and (\ref{ct3}) define the new strings, loop operators in terms of the old Kogut-Susskind link operators. These relations, being canonical, can also be inverted. They will enable us  
to rewrite Kogut-Susskind Hamiltonian in terms of loop operators in section 5.  The loop flux operators are trivial to invert and we get: 
\begin{align}
 &U(1) = {\cal U}_1,~~~ ~~~~~~~~U(2) = {\cal U}_1^\dagger {\cal U}_2,  \nonumber\\ 
 &U(3) ={\cal U}_2^\dagger {\cal U}_3,~~~ ~~~~~
 U(4) = {\cal U}_3^\dagger {\cal W}.    
\label{irc} 
\end{align}
The electric field relations are also easy to invert\footnote{
We have used canonical electric field relations (\ref{ct1}), (\ref{ct2}) and (\ref{ct3}) to derive (\ref{ir}). As an example:
\begin{align} 
E_-^a(1) \equiv -R_{ab}(U_1) E_+^b(1) = -R_{ab}(U_1)\left[{\cal E}_+^b(1)-E_-^b(2)\right]=   {\cal E}_-^a(1) ~~~~~~\nonumber \\
- R_{ab}({\cal U}_2)E_+^b(2) ={\cal E}_-^a(1)- R_{ab}({\cal U}_2) \left[{\cal E}^b_+(2) - E_-^b(3)\right]      
={\cal E}_-^a(1) + {\cal E}_-^a(2) \nonumber \\
-R_{ab}({\cal U}_3) E^b_+(3)= {\cal E}_-^a(1) + {\cal E}_-^a(2) -R_{ab}({\cal U}_3)\left[{\cal E}_+^b(3) - E_-^b(4)\right]=  {\cal E}_-^a(1) 
\nonumber \\ 
+ {\cal E}_-^a(2)+{\cal E}_-^a(3)  -R_{ab}({\cal W}) E^b_+(4) = {\cal E}_-^a(1) + {\cal E}_-^a(2) +{\cal E}_-^a(3) +{\cal E}_-^a. ~~~\nonumber
\end{align}
Similarly the other three relations in (\ref{ir}) can be derived.}:
\bea 
E^a_-(1) &= &{\cal E}^a_-(1)+ {\cal E}^a_-(2)+ {\cal E}^a_-(3)+{\cal E}^a_-, \nonumber \\
E^a_-(2) &= &R_{ab}({\cal U}^\dagger_1) ~\big[{\cal E}^b_-(2)+ {\cal E}^b_-(3)+{\cal E}^b_-\big], \nonumber \\ 
E^a_-(3) &= &R_{ab}({\cal U}^\dagger_2) ~\big[{\cal E}^b_-(3)+{\cal E}^b_-\big], \nonumber \\ 
E^a_-(4) &= &R_{ab}({\cal U}^\dagger_3) ~{\cal E}^b_-. 
\label{ir}
\eea
Note that (a) the relations (\ref{irc}), ~(\ref{ir}) are consistent with the gauge transformation properties of
the Kogut-Susskind link operators and the newly defined loop operators, (b) As expected, the inverse relations (\ref{ir}) transform the original Gauss law in terms of 
link operators at the origin, 
$E_-^a(1) + E_+^a(4) =0,$
into the Gauss law in terms of the loop 
operators\footnote{We have used  
$E_-^a(1) \approxeq {\cal E}^a$ as ${\cal E}(I) = 0, 
~I=1,2,3$ in ${\cal H}^p$ and 
$E_+^a(4) \equiv -R_{ab}(U^\dagger(4)) E_-^a(4) = 
-R_{ab}({\cal W^\dagger}) {\cal E}_-^b = {\cal E}_+$.} 
in (\ref{gl}), ${\cal E}_-^a+{\cal E}_+^a=0$,   
 in ${\cal H}^p$,  (c) the gauge invariant operators, e.g., the quadratic electric field terms (see Kogut-Susskind Hamiltonian (\ref{ks}) in section 5),  do not contain unphysical  string variables. 
Further,  we note that:
\begin{itemize} 
\item  Every lattice site (x,y) away from the origin 
(the corners A, B, C in this simple case) is attached  with a  string ${\cal U}(x,y)$ ~(${\cal U}_1, {\cal U}_2, { \cal U}_3$ in our case). They all start from the origin. 
\item  
The Gauss law operators at these sites  
are the right electric 
fields of the strings. In the single plaquette case:   
\begin{align}  
{\cal G}^a(1,0) ={\cal E}^a_+(1),  
{\cal G}^a(1,1) = {\cal E}^a_+(2), 
{\cal G}^a(0,1) ={\cal E}^a_+(3), 
\nonumber 
\end{align}       
making these strings devoid of any physical consequences. 
\end{itemize}
The above features of the canonical transformations,  
converting links into physical loops and unphysical  strings, are general and valid for any SU(N) in any 
space  dimension.  
\begin{figure}
\centering
\includegraphics[scale=1.4]{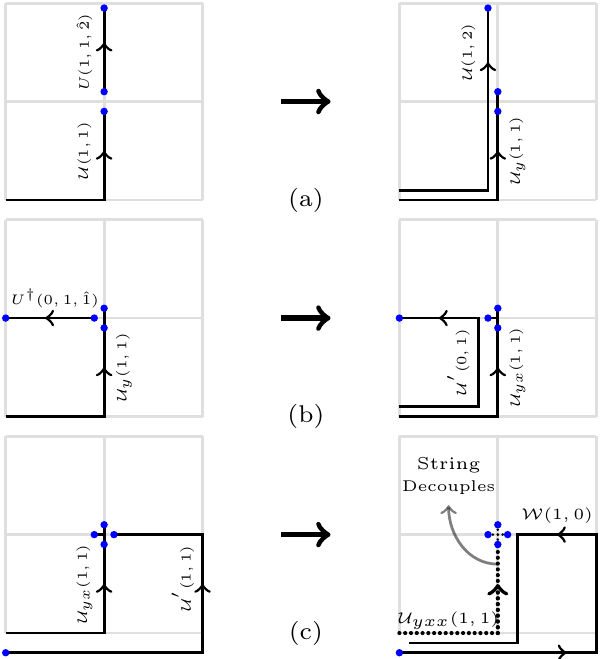}
\caption{Formation of plaquette loops and the decoupling of strings. The loop plaquette operator ${\cal W}(1,0)$ is shown.  The ${\textcolor{blue} \bullet}$  represent the electric fields of the links, strings and loops. In the last figure, the right electric field of the string operator ${\cal U}_{yxx}(1,1)$
is the Gauss law at (1,1): ${\cal U}_{yxx+}^a(1,1) = 
{\cal G}^a(1,1)$.}
\label{upso}
\end{figure} 

\subsection{${\cal P}$ Plaquettes case}
On a $d=2$ lattice with open boundary conditions, the canonical transformations transform ~${\cal L}$ ~Kogut-Susskind link operators into ${\cal N}-1$ unphysical strings operator
and ${\cal P}$ physical loop operators. The degrees of freedom match as: ${\cal L} = {\cal P} + ({\cal N}-1)$.   
The $({\cal N} -1)$ strings ${\cal U}(x,y)$, attached to the lattice site $(x,y)$ away from the origin, are along the oriented paths:~$(0,0)\rightarrow (x,0) \rightarrow (x,y)$. They are unphysical and decouple as a consequence of Gauss laws. As an example,  we show in figure (\ref{upso}) how the final string attached to the lattice site $(1,1)$ decouples. In figure (\ref{upso}-a), we attach the Kogut-Susskind link operator $U(1,1;\hat 2)$ to the string ${\cal U}(1,1)$ to obtain the new strings ${\cal U}_y(1,1)$ and ${\cal U}(1,2)$. 
The subscript y of ${\cal U}_y(1,1)$ shows that it has been extended in the y direction and its right electric field at $(1,1)$ is the sum of two Kogut-Susskind electric fields in y directions. These two Kogut-Susskind electric fields are shown by the two blue dots at $(1,1)$ in the right hand side of figure (\ref{upso}-a).
We now extend the string ${\cal U}_y(1,1)$ in the -ve x direction by attaching it with the Kogut-Susskind link operator $U^\dagger(0,1,\hat {1})$ as shown in figure (\ref{upso}-b). This results in new strings ${\cal U}_{yx}(1,1)$ and 
${\cal U}^\prime(0,1)$ as shown in the right hand side of figure ({\ref{upso}-b). Now the right electric field of ${\cal U}_{yx}(1,1)$ at $(1,1)$ is the sum of three Kogut-Susskind electric fields at this point as shown by the three blue dots in the right hand side of figure (\ref{upso}-b). We finally extend ${\cal U}_{yx}(1,1)$ in the +ve x direction by attaching it with $U^\prime(1,1)$ as shown in the left hand side of figure (\ref{upso}-c). The result is   ${\cal U}_{yxx}(1,1)$ and the plaquette loop operator ${\cal W}(1,0)$. The right electric field of ${\cal U}_{yxx}(1,1)$ is now sum of all four Kogut-Susskind electric fields at $(1,1)$ as shown by the four blue dots in the right hand side of figure (\ref{upso}-c).  The Gauss law implies that the string ${\cal U}_{yxx}(1,1)$ excitations are outside ${\cal H}^p$.  Therefore it can be ignored. 

From now onwards,  we focus on the remaining physical ${\cal P}$ loop operators  
denoted by ${\cal W}(x,y)$. They are along the paths: $(0,0) \rightarrow (x,0) \rightarrow (x,y) \rightarrow (x+1,y) \rightarrow (x+1,y+1) \rightarrow (x,y+1) \rightarrow (x,0) \rightarrow (0,0)$.  These paths are shown in figure (\ref{f:4}) for a small lattice with $N=2$. It is convenient to collectively denote them by ${\cal W}(p), ~p=1,2, \cdots ,{\cal P}$.  The physical SU(2) prepotentials are defined as: 
\begin{align}  
\label{holop}
{\cal E}_-^a(p) = \frac{1}{2}a^\dagger(p)~\sigma^a~ a(p),~~~{\cal E}_+^a(p) = \frac{1}{2} b^\dagger(p)~ {\sigma^a}  ~b(p) ~~~~~~~~ \\ 
{\cal W}_{\alpha\beta}(p) =  \frac{1}{\sqrt{(N(p)+1)}} \Big(
{\cal W}_{\alpha\beta}^+(p) 
 - 
{\cal W}_{\alpha\beta}^{-}(p)\Big) 
\frac{1}{\sqrt{(N(p)+1)}}. \nonumber     
\end{align}
In (\ref{holop}), ${\cal W}^+_{\alpha\beta}(p) \equiv 
\tilde a^{\dagger}_{\alpha}(p)~  b^{\dagger}_{\beta}(p)$
and ${\cal W}^-_{\alpha\beta}(p) \equiv a_{\alpha}(p) ~\tilde b_{\beta}(p)$.  All  loop operators in (\ref{holop}) are invariant under gauge transformations anywhere on the lattice except origin. Under gauge transformations at the origin, they all transform together like adjoint matter fields: 
\begin{align} 
\label{gtl} 
&{\cal E}_{\mp}(p) \rightarrow \Lambda_0~ {\cal E}_{\mp}(p) ~\Lambda^\dagger_0, ~~~~~~~~p=1,2,\cdots {\cal P}, 
\nonumber \\ 
&{\cal W}(p)  \rightarrow \Lambda_0 ~{\cal W}(p)  ~\Lambda^\dagger_0; ~~~~~~~~~p=1,2,\cdots {\cal P}.   
\end{align}
 All $2{\cal P}$ prepotential creation operators transform 
as matter doublets: 
\begin{align} 
\label{gtlpp} 
&{a}_\alpha (p) \rightarrow \left(\Lambda_0\right)_{\alpha\beta} {a}_\beta (p), ~~~ {b}_\alpha (p) \rightarrow \left(\Lambda_0\right)_{\alpha\beta} {b}_\beta (p).   
\end{align} 
The Gauss law at the origin reduces to simple constraints:
 \begin{align} 
{\mathcal G}^a = \sum_{p=1}^{P}  {\sf L}^a(p) =  \sum_{p=1}^{P}  \Big({\cal E}_-^a(p) + {\cal E}_+^a(p)\Big) \equiv {\sf L}^a_{total}  \approxeq 0.
\label{glpp}
\end{align} 
This is a straightforward generalization of the single plaquette result (\ref{gtlp}).
Thus after these canonical transformations, the nontrivial issue of non-abelian gauge invariance 
reduces to a much simpler issue of global invariance.
We solve these constraints in the next section 
to construct a basis in ${\cal H}^p$.

\begin{figure}[t]
\begin{center}
\includegraphics[scale=.65]
{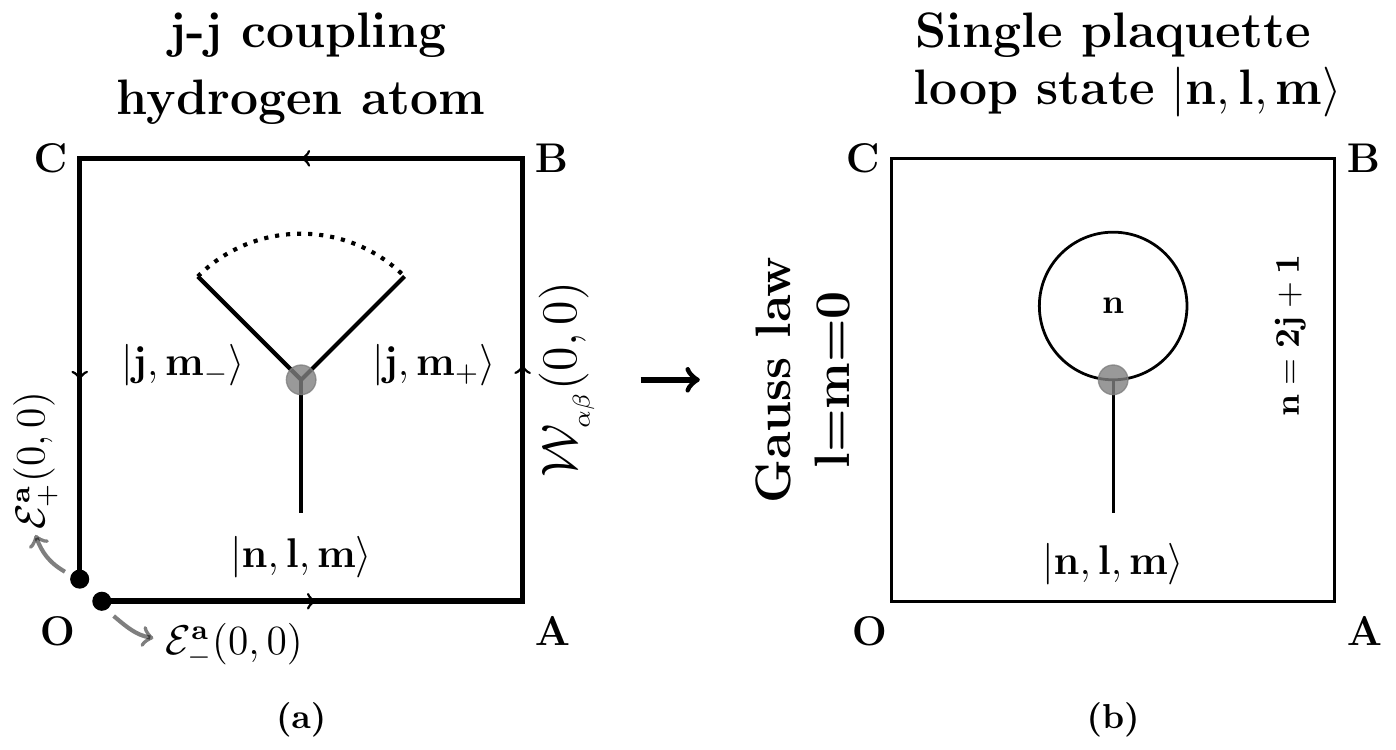} 
\end{center}
\caption{A graphical tadpole representation  of hydrogen atom states $|n~l~m\rangle$ 
or equivalently a SU(2) loop state over  a plaquette.
The dotted arch represents $j_+=j_-=j$ in the j-j coupling (\ref{nlm}) denoted by \textcolor{gray}{$\bullet$}. The tadpole loop  represents the SU(2) flux circulating  within a plaquette.  The vertical leg of tadpole represents leakage of angular momentum flux $(l,m)$ through the plaquette.}
\label{fama}
\end{figure} 

\section{Loop States \& Hydrogen Atoms}
\label{slsha}
We first start with the simple single plaquette 
case. 
We identify the hydrogen atom angular momentum, Lenz vector  operators with the SU(2) 
loop electric field operators of the gauge theory:
$$J^a_{\mp} ~\leftrightarrow ~ {\cal E}_\mp^a  
.$$
This identification further implies:~$L^a \leftrightarrow {\sf L}^a$ where $L^a$ are the angular momentum operators of a hydrogen atom and ${\sf L}^a$ are the generators of Gauss law (\ref{gl}) in a single plaquette case.  
This immediately implies that  
CSCO-I and CSCO-II  of hydrogen atom also characterize lattice gauge theory Hilbert space. 
The CSCO-II is even more natural for gauge theory as the remaining three 
SU(2) Gauss law constraints at the origin (\ref{gl}) are 
trivially removed in this coupled basis. In the single 
plaquette case,  the non-abelian Gauss laws  (\ref{gl}) state that the spherically symmetric hydrogen atom states $|n ~l=0~m=0\rangle$ form an orthonormal loop basis in ${\cal H}^{P}$:
\begin{align} 
|n\rangle \equiv |n, l=0, m=0\rangle & = \frac{\left(k_+\right)^n}{\sqrt{n!(n+1)!}} ~|0\rangle. \nonumber \\
\sum_{n=1}^{\infty} ~|n\rangle\langle n| = {\cal I}, ~~~~~&~~\langle m|n\rangle = \delta_{nm}.
\label{spls} 
\end{align} 
In (\ref{spls}) ${\cal I}$ is the identity operator in ${\cal H}^p$. The loop creation-annihilation and number operators
are defined as:  $k_+ \equiv a^\dagger \cdot \tilde b^\dagger, ~~k_- \equiv a\cdot \tilde b$ and $
k_0 \equiv 1/2(N_a+N_b+2)$. They satisfy SU(1,1) 
algebra: $[k_-,k_+] = 2 k_0, ~ [k_0, k_\pm] = \pm k_\pm$. 

 At this stage, before going to ${\cal P}$ plaquette case,  it is convenient to graphically represent the coupled state $|n~l~m\rangle$ in (\ref{nlm}) on a 
plaquette by a tadpole diagram as shown in figure (\ref{fama}-a,b).
The  loop at the top of the tadpole in a plaquette represents the non-abelian flux circulating in a loop within the  plaquette. 
The vertical leg of the tadpole, on the other hand,  represents the leakage of 
non-abelian flux through the plaquette. More precisely,  the anti-symmetric part  of the coupling between $a^\dagger$ and $b^\dagger$ (Young tableau boxes arranged vertically in pairs)
present in $|j,m_-\rangle$ and  $|j,m_+\rangle$ states is shown by the circle. The  leg represents the symmetric part of this coupling  (Young tableau boxes arranged along a row).
We now draw  tadpoles over each of 
the ${\cal P}$ plaquettes and then couple their emerging  angular momentum fluxes $(l_p,m_p)$ with $p=1,\cdots, {\cal P}$  in a sequential manner as in figure (\ref{f:4}-b). It corresponds to 
going from decoupled tadpole basis diagonalizing $3{\cal P}$ CSCO-II operators $\Big\{J^2_p,~L^2_p,~ L^{a=3}_p\Big\}$, with eigenvalues  $(j_p(j_p+1),~l_p(l_p+1), ~m_p)$, to a coupled basis which diagonalizes the following coupled angular momentum operators:
\begin{align} 
\left\{ \begin{array}{ccccc} \vec J^{~2}_1 & \vec J^{~2}_2 &  \cdots \vec J^{~2}_{{\cal P}-1}& \vec J^{~2}_P \\ 
\vec L^2_1 & \vec L^2_2 &  \cdots  \vec L^2_{{\cal P}-1} & \vec L^2_{\cal P} \\ 
(\vec L_{12})^2 & (\vec L_{123})^3& \cdots {(\underbrace{~\vec L_{total}~}_{=0})^2} & \underbrace{\vec L^{a=3}_{total}}_{=0}\ \end{array} \right\}. \nonumber 
\end{align} 
Above $(\vec L_{1,2,\cdot q})^2 \equiv (\vec L_1+\vec L_2+ \cdots \vec L_q)^2$ with eigenvalue $l_{1,2\cdots q}(l_{1,2\cdots q}+1)$ and $q=2,3, \cdots, {\cal P}$. We have put $\vec L_{total} =0$ because of the Gauss law (\ref{glpp}). This further implies 
that $l_{1,2, \cdots {\cal P}-1} = l_{\cal P}$ as $\vec L_{total} \equiv  \left(\vec L_{1,2, \cdots {\cal P}-1}+ \vec L_{\cal P} \right)=0$.
The resulting orthonormal and complete loop basis
is: 
\begin{align}
\left\vert \begin{array}{cccc} n_1 & n_2 &  ~~\cdots ~~~n_P ~~\\ 
l_1 & l_2 &  \cdots   ~~~l_P \\ 
l_{12} & l_{123}& ~~~~~\cdots  l_{12\cdots {\cal P}-2} \end{array} \right\rangle \equiv  ~~~~~~~~~~~\nonumber \\
\Big\{\underbrace{|n_1~l_1~m_1\rangle \otimes  |n_2~l_2~m_2\rangle 
\cdots \otimes |n_P~l_P~m_P\rangle}_{\bf{Wigner~ coupled~ states~of~hydrogen~atoms}}\Big\}^{{l_{total}=0}}_{{m_{total}=0}} 
\label{entlat} 
\end{align}

\begin{figure}[t]
\centering
\includegraphics[scale=.55]
{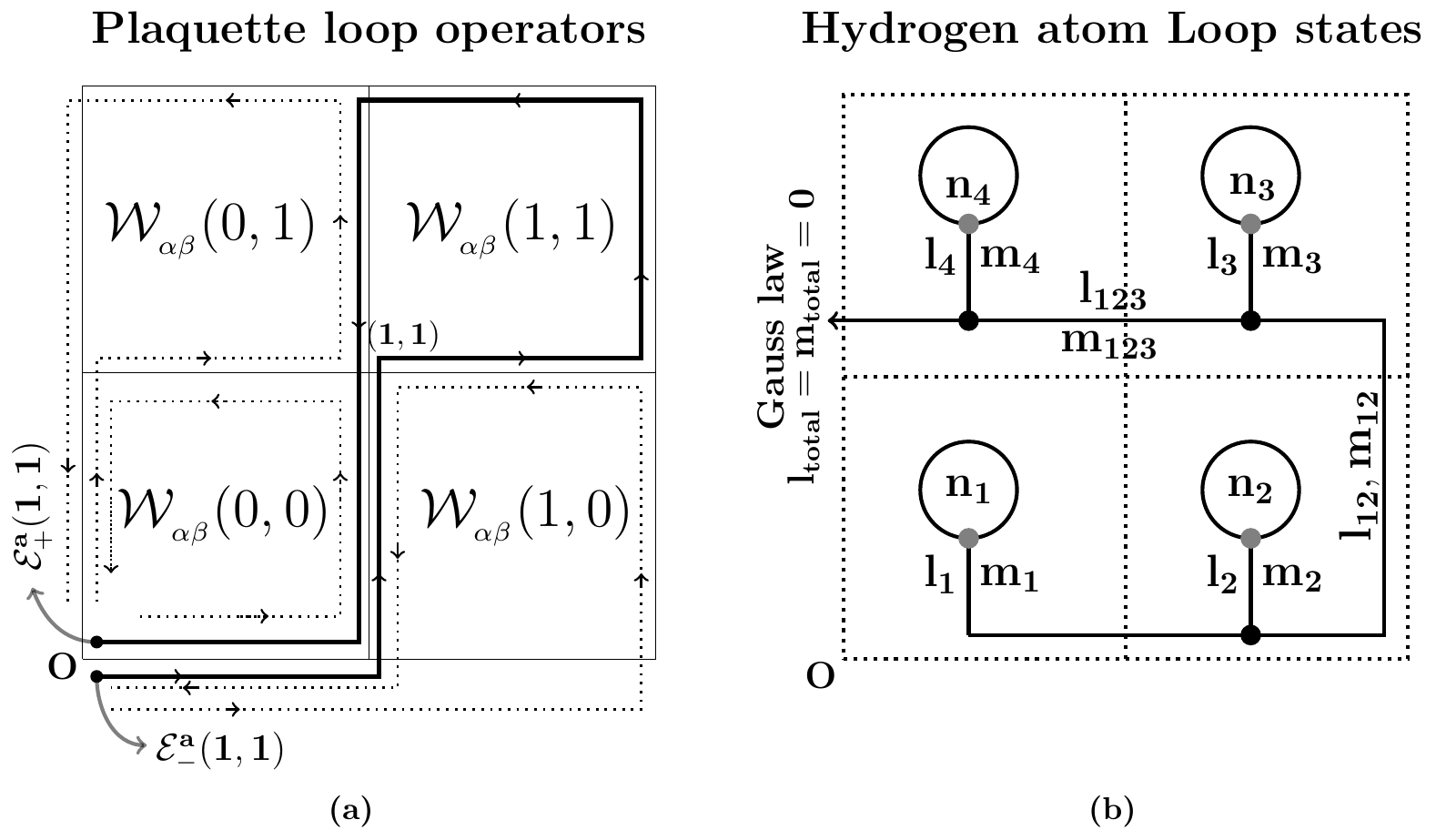} 
\caption{[a] ~The construction of the plaquette loop operators ${\cal W}_{\alpha\beta}(x,y)$ and ${\cal E}^{a}_{\mp}(x,y)$
in a  four plaquette case. [b]~The SU(2) hydrogen atom/tadpole basis with  hydrogen atom  states $|n~l~m\rangle$. 
The \textcolor{gray}{$\bullet$} and $\bullet$ in [b] represent $jj$  and $ll$ couplings in (\ref{nlm}) and (\ref{exc}) respectively.} 
\label{f:4}
\end{figure} 
The above loop basis (shown in figure (\ref{f:4}-b) for 4 plaquette case) will be briefly denoted\footnote{ 
As an example the loop states over 4 plaquettes in figure (\ref{f:4}-b) are constructed as; 
\begin{align} 
|[n]~[l]~[ll] \rangle &= \sum_{m_1m_2m_3m_4m_{12}}
C_{l_1m_1,l_2m_2}^{~l_{12}m_{12}} C_{l_{12}m_{12},l_3m_3}^{~l_{123}m_{123}} C_{l_{123}m_{123},l_4m_4}^{~~~0~0} \nonumber \\
&|n_1~l_1~m_1\rangle~ ~|n_2~l_2~m_2\rangle~|n_3~l_3~m_3\rangle~
|n_4~l_4~m_4\rangle.
\label{exc} 
\end{align} 
The states $|n~l~m\rangle$ are constructed in (\ref{csco1}), (\ref{nlm}) and figure (\ref{fama}).
They  are orthonormal as well as complete: 
\begin{align} 
\langle ~[\bar n] ~[\bar l]~[\bar{ll}]~|~[n]~[l]~[ll]~\rangle  = \delta_{[\bar n], [n]} ~\delta_{[\bar l],[l]}~~\delta_{[\bar{ll}],[ll]} \\ 
\sum_{[n]}\sum_{[l]}\sum_{[ll]}  ~|~ [ n] ~[ l]~[{ll}]~\rangle ~\langle ~[n]~[l]~[ll]~| = {\cal I}.
\label{bbbb}  
\end{align}  
The operator ${\cal I}$ denotes the identity operator in ${\cal H}^p$.}   by $\left\vert{[n]~ [l]~  [ll]}\right\rangle$.
The symbols $[n];[l]$ and $[ll]$ stand for the sets $(n_1 \cdots n_{\cal P})$: ${\cal P}$ principle quantum 
numbers;~ 
$[l_1\cdots l_{\cal P}]$: ${\cal P}$ angular momentum quantum numbers 
and  $(l_{12},l_{123}, \cdots ,l_{123 \cdots ({\cal P}-2)})$: $({\cal P}-3)$ coupled angular momentum quantum numbers respectively.  
Thus the hydrogen atom loop basis (\ref{entlat}) 
in ${\cal H}^p$ is labelled by ${\mathbb N} = 3({\cal P}-1)$ gauge invariant quantum numbers. As expected, this is also the dimension of quotient space ${\mathbb N} = \left[\frac{\otimes_{links}SU(2)}{\otimes_{sites}SU(2)}\right]$ in $d=2$. 

We now analyze this equivalence in the  dual magnetic description. 
We again start with single plaquette basis $|j~m_-~m_+\rangle$  in (\ref{csco1}) and  make a duality  transformation to define states on SU(2) group manifold $S^3$ as: 
\begin{align}
\vert \Omega_W \rangle = 
\sum_{j=0}^{\infty}\sum_{m_{\mp} =-j}^{+j} 
\{j\} 
~  D^{~j}_{ m_- m_+}(\Omega_W)~ 
 \vert j, m_-, m_+\rangle 
\label{ft} 
\end{align}
In (\ref{ft}) $\{j\} \equiv (2j+1)^{\frac{1}{2}}$,~  $D^{~j}_{m_- m_+}(\Omega_W)$ are the Wigner matrices 
characterized by SU(2) group manifold $S^3$: 
\bea 
{\Omega}_{{W}}\left(w_0,\vec w\right) \equiv w_0 ~\sigma_0 + i \vec w \cdot \vec \sigma, ~~~  w_0^2+\vec w^2 =1: S^3.   \nonumber 
\label{s3} 
\eea 
The recursion relations of Wigner matrices show \cite{ms} 
that the orthonormal and complete angular states (\ref{ft}) also diagonalize the plaquette 
loop operators ${\cal W}_{\alpha\beta}$ in (\ref{holo}): 
\bea 
{\cal W}_{\alpha\beta} ~ \vert {\Omega_W}  \rangle = 
\big({\Omega_W} \left(\omega_0, \vec w\right) \big)_{\alpha\beta} ~  \vert {\Omega_W}  \rangle.
\label{wgt}
\eea 
Under global SU(2) transformation (\ref{gtl}):
\bea 
\vert {\Omega_W } \rangle \rightarrow \vert \Lambda_o ~{\Omega_W}  ~\Lambda_o^\dagger \rangle.
\label{angletrans}
\eea 
The gauge generators $\sf L^1, ~\sf L^2, ~\sf L^3$ 
in (\ref{gl}) rotate $(w_2w_3),~(w_3w_1),~(w_1w_2)$ 
planes respectively leaving $w_0$ (gauge) invariant. 
Defining ``Lenz operators" 
in lattice gauge theory as ${\sf A}^a \equiv  {\cal E}_+^a-{\cal E}_-^a$, we see that $({\sf A}^1,{\sf A}^2, {\sf A}^3)$
generate rotations in $(w_0w_1),~(w_0w_2),~(w_0w_3)$ 
planes respectively. Therefore, the actions of $({\sf L^a},{\sf A^a})$ on $\Omega_W$ in gauge theory is exactly same as the actions of  $(L^a,A^a)$ on $\Omega_H$  in hydrogen atom. Therefore, we 
further identify:
$$ 
\Omega_H  \sim \Omega_W \equiv \Omega. $$

\section{SU(2) Loop Dynamics \& Loop Operators}
\label{sldlo}
We now construct non-abelian loop dynamics directly in terms of the loop operators ${\cal W}_{\alpha\beta}(x,y)$ and their conjugate loop electric fields ${\cal E}^a_\mp(x,y)$ constructed in section (\ref{sct}).  We start with the Kogut Susskind Hamiltonian \cite{kogut}:  
\begin{align}
H = g^2\sum_{l}  \vec E^{2}_l + \frac{K}{g^2} 
\sum_{p} \big(2 - Tr ~U_{p}\big). 
\label{ks} 
\end{align}
In (\ref{ks}) K is a constant, $l \equiv (x,y,\hat i)$ 
denotes a link in $\hat i$ direction, $p$ denotes a plaquette and 
$Tr~ U_p \equiv Tr~\left(U_1U_2U_3^\dagger U_4^\dagger\right)$
is the  magnetic field term on a plaquette  $p$. In the simple single plaquette case, the Kogut-Susskind Hamiltonian (\ref{ks}) in terms of the loop operators is:  
\begin{align} 
H =  4 g^2 {\cal E}^2 - 
\frac{K}{g^2} 
\big(2 -Tr ~{\cal W}\big). 
\label{ham} 
\end{align}
In going from link description (\ref{ks}) to loop description(\ref{ham}) we have used (\ref{ct3}) and (\ref{ir}). The first term $4 \vec {\cal E}^2 (\equiv 4 \vec {\cal E}_-^2
= 4 \vec {\cal E}_+^2)$ describes  the loop electric fields. The factor 4 is because of the 4 links on the plaquette and 
the relations 
(\ref{ir}) after
 putting  the three string electric fields ${\cal E}^a_\mp(1) =
 {\cal E}^a_\mp(2) = {\cal E}^a_\mp(3) \approxeq 0$.
The Hamiltonian (\ref{ham}) is invariant under global SU(2) 
transformations (\ref{gtlp}). 

Similarly,  the Kogut Susskind 
Hamiltonian in the loop space on a finite lattice is \cite{ms}: 
\begin{align} 
 H  =  \sum_{(x,y)} \Bigg\{ g^2 {\vec E}^2(x,y,\hat 1)+ g^2 {\vec E^2(x,y,\hat 2)}   ~~\nonumber \\ 
+ ~
\frac{K}{g^2}~ 
\Big(2 - Tr~ {\cal W}(x,y)\Big)\Bigg\}. 
\label{loopham} 
\end{align} 
In (\ref{loopham}), $E^2(x,y,\hat{i})=E_-^2(x,y,\hat{i})=E_+^2(x,y,\hat{i})$ as discussed in section 3 and  we have used the property of the canonical transformations: $Tr~U_p(x,y) 
= Tr \left({\cal U}^\dagger(x,y){\cal W} (x,y){\cal U}(x,y)\right) = Tr{\cal W}(x,y),$ for  converting  the 
magnetic field term.  The Kogut-Susskind electric fields above are given by \cite{ms}: 
\begin{align}
E_-^a(x,y,\hat 1) \approxeq& R_{ab}({\cal U}(x,y))\Big[{\cal E}^a_-(x,y) + {\cal E}^a_+(x,y-1) + {\Delta}^a_{X}\Big]  \nonumber \\
E^a_-(x,y,\hat 2) \approxeq& R_{ab}({\cal U}(x,y)) \Big[{\cal E}^b_+(x,y) +    \nonumber\\   
& ~~~~~~~~~~~~~~~~~{\tiny {\cal R}_{_{\cal W}}(x,y)}_{bc}{\cal E}^c_-(x-1,y) + \Delta_Y^b   ~\Big]
\label{irlp} 
\end{align} 
In (\ref{irlp}), $E_-^a(x,y,\hat{i})$ is the electric field at $(x,y)$ in the $\hat{i}$ direction and rotates the link operator $U(x,y,\hat{i})$ from the left as in (\ref{ccr}). $\Delta_{X}, \Delta_Y$ and ${\cal R}_{\cal W}$ 
are defined as: 
\begin{align}
{\Delta}^a_{X}(x,y)  \equiv  \delta_{y,0} \sum_{r=x+1}^{{\cal N}}\sum_{s=0}^{{\cal N}} {\sf L}^a(r,s),   
~~{\Delta}^a_{Y}(x,y) ~~ ~~~ \nonumber\\ 
 \equiv   \sum_{s=(y+1)}^{{\cal N}} {\sf L}^a(x,s),~ R_{_{\cal W}}(x,y) \equiv  \prod_{q=0}^{y-1} R\big({\cal W}(x-1,q)\big) 
\label{nlt}  
\end{align}
Note that the physical  loop operators ${\cal E}^a_{\mp}(x,y), ~{\cal W}(x,y)$ are located at $(x,y)$ as shown in figure (\ref{f:4}-a). There are no gauge fields.  
The relations (\ref{irlp}) are generalizations of the simple  single plaquette results in (\ref{ir}).  Like in single plaquette case these relations are arrived at by  systematically inverting the canonical transformations at each stage through out the lattice \cite{ms}.   
Again, as expected, the unphysical strings ${\cal U}(x,y)$ attached to sites (x,y) in (\ref{irlp})  disappear from the Hamiltonian (\ref{loopham}) as $R^TR=1$. 
 
The Hamiltonian ({\ref{loopham}) describes dynamics 
directly in terms of the  essential loop creation, annihilation operators and their conjugate electric 
fields. There is no local gauge invariance and no 
gauge fields.  The gauge transformations at the origin manifest themselves in the global SU(2) invariance (\ref{gtl}) of the Hamiltonian (\ref{loopham}).

We note that in going from links to loops ((\ref{ks}) to (\ref{loopham})),  all interactions have shifted 
from the ($1/g^2$) magnetic field term to ($g^2$) 
electric field term. Therefore, as opposed to strong coupling  ($g^2 \rightarrow \infty$) expansion 
with simple $g^2 E^2$ term in (\ref{ks}), the 
loop formulation (\ref{loopham}) with the simple $1/g^2~ Tr{\cal W}$ term provides an alternative  framework 
appropriate to develop gauge invariant weak coupling ($g^2\rightarrow 0$) expansion near the continuum limit\footnote{In strong coupling ($g^2 \rightarrow \infty$) expansion non-interacting terms $g^2E^2_{link}$ are trivially diagonalized and 4 flux interaction terms $1/g^2~ Tr(U_1U_2U^\dagger_3U^\dagger_4)$ are  treated in perturbation. However, one is far away from continuum.}.
In fact, the magnetic field term ${1}/{g^2} ~Tr~{\cal W} \sim 1/g^2~ \left(k_++k_-\right)$ dominating in this limit,  has the simplest possible action on the hydrogen atom loop basis (\ref{entlat}). Using (\ref{holop}) and (\ref{entlat}), the matrix elements are: 
\begin{align} 
\langle~ [\bar n] ~[\bar l]~[\bar{ll}]~ | ~Tr {\cal W}(p)~ |~[n]~[l]~[ll]~\rangle    ~~~~~~~
  \nonumber\\  
= \delta_{[\bar l],[l]}~~\delta_{[\bar{ll}],[ll]}\delta_{[\bar n]^\prime,[n]^\prime}~\Big[\delta_{\bar n_p, n_p+ 1} +  \delta_{\bar n_p, n_p- 1}\Big].   
\label{metw} 
\end{align}
Above $n_p = 2j_p+1$ and $\bar n_p \equiv 2\bar j_p +1$
and $\delta_{[\bar n]^\prime,[n]^\prime}$ denotes the delta function  over $({\cal P}-1)$ principal quantum numbers over all plaquettes $p^\prime\neq p$. 
This should be compared with  the  action of the standard magnetic field term  $1/g^2~Tr~ (U_1U_2U^\dagger_3U^\dagger_4)$  on the standard SU(2) spin network basis describing orthonormal loop states \cite{manu}. This action is extremely complicated and the matrix elements are  given by 18-j (30-j) Wigner coefficients in $d=2 ~(d=3)$ dimension as compared to the 
simple   $\delta$ function over a single principal quantum number in (\ref{metw}).

%
All interactions in the Hamiltonian (\ref{loopham}) in this loop formulation are contained in (\ref{irlp}). They   describe nearest neighbour loop-loop interactions  through their electric fields along with non-local terms $\Delta_X(x,y),~\Delta_Y(x,y),~{\cal R}_{_{\cal W}}(x,y)$. However, these non-local interactions can be ignored at tree level  
  near the continuum ($g^2\rightarrow 0$) limit as in this limit ${\sf L}^a(x,y) \equiv {\cal E}^a_+(x,y) + {\cal E}^a_-(x,y) \rightarrow 0$ and 
  $R_{ab}({\cal W})(x,y) \rightarrow \delta_{ab}$. Thus we 
  have a  generalized ``spin model" with nearest neighbour interactions along with spin creation ($k_+$) and spin annihilation ($k_-$) terms.
   The analysis of this model through variational method and coupled cluster expansion is  under study and will be reported 
  elsewhere.
  
  \section{Summary \& Discussion}
  \label{ssnd} 
  
  In this work we have obtained a loop formulation of 
  pure SU(2) lattice gauge theory through a series of canonical transformations. These canonical transformations, constructed in section (\ref{sct}), are 
  also valid  for any SU(N) group. They  can also  be extended to higher dimension. As shown in this work the problem of over-completeness of SU(N) loop states leading to SU(N) Mandelstam  constraints amongst Wilson loops is bypassed in this approach.
   To the best of our knowledge the solutions of SU(N) 
   Mandelstam constraints for $N \ge 3$ do not exist 
   \cite{loops}. The SU(N) canonical transformations
   discussed in this work also provide a complete set of mutually independent SU(N) Wilson loop operators.  
   
  In the case of SU(2), a gauge invariant state can be expanded in the hydrogen atom basis. Therefore, any gauge invariant operator acting on a physical state generates transitions in 
    the underlying basis states $|n~l~m\rangle$. Such transitions are produced by the generators of the 
    dynamical symmetry group SO(4,2) of hydrogen atom. 
    Therefore, SU(2) loop Hamiltonian can also be described in terms of SO(4,2) generators. These results  will be reported later. 
  
  Having removed all redundant gauge and loop degrees of freedom through canonical transformations,  we should further identify the relevant and interesting part of ${\cal H}^p$ for low energy physics. This can be done using tensor networks or matrix product ideas \cite{tna} as the 
  $3({\cal P}-1)$ gauge invariant quantum numbers can 
  be thought of as generalized spins over plaquettes of the lattice. 
 In context of cold atoms, imposing non-trivial and exotic non-abelian Gauss law constraints experimentally at every lattice site  is a challenging task. Different ways of dealing with these local constraints have been proposed in 
 the past \cite{stan}.  The present formulation 
 with hydrogen atoms naturally solve these constraints
 and may be useful in the cold atom experiments in the future.  
This  absence of local Gauss laws should  also help us define entanglement entropy of two complimentary regions in a gauge invariant state bypassing obstacles \cite{ee} created by them at the boundary of the two regions. The  two regions can have mutually independent hydrogen atom/tadpole basis which are joined together  across the boundary at the end. 
%


\vspace{0.4cm}

\noindent {\it Acknowledgments: MM thanks  Ramesh Anishetty  and H. S. Sharatchandra 
for useful discussions. MM especially acknowledges  H S Sharatchandra 
for introducing him to the canonical transformations which  were crucial for this work and for initial discussions on the resulting  Hamiltonian.
TPS thanks CSIR for financial support.}

\end{document}